\documentclass[preprint,prb]{revtex4}

\usepackage{amsmath}
\usepackage{amssymb}

\newcommand{\bff}{\boldsymbol{f}}

\newcommand{\gradv}{\boldsymbol{\nabla}}
\def\v#1{{\bf#1}}

\begin{document}

\title{A formal interpretation of the displacement current and the instantaneous formulation of Maxwell's equations}
\author{Jos\'e A. Heras}
\email{herasgomez@gmail.com}
\affiliation{Universidad Aut\'onoma Metropolitana Unidad Azcapotzalco, Av. San Pablo No.\ 180, Col. Reynosa, 02200, M\'exico D. F. M\'exico}

\begin{abstract}
Maxwell's displacement current has been the subject of controversy for more than a century. Questions on whether the displacement current represents a true current like the conduction current and whether it produces a magnetic field have recently been discussed in the literature. Similar interpretations for the Faraday induction current have also been controversial. These basic questions are answered in this paper by considering the relation between the displacement and conduction currents as well as the relation between Faraday induction and conduction currents. It is pointed out  that the displacement current contributes to the magnetic field and that the induction current contributes to the electric field.
However, the displacement and induction currents cannot be considered as the conduction current because they are nonlocal. Both relations are used to implement an instantaneous formulation of Maxwell's equations with local and nonlocal sources.
\end{abstract}

\maketitle

\section{Introduction}
Maxwell's displacement current $\epsilon_0\partial \v E/\partial t$ has elicited long-standing controversies regarding its interpretation as a true source. The controversy began when Maxwell used the phrase ``the true electric current" to refer to the sum of the conduction current and the displacement current: $\v J+\epsilon_0\partial \v E/\partial t$, which suggests that both currents have the same level of reality. Maxwell considered the current $\epsilon_0\partial \v E/\partial t$ as being ``electromagnetically equivalent" to the current $\v J$. Questions on whether $\epsilon_0\partial \v E/\partial t$ is a true current like $\v J$, and whether it produces a magnetic field have recently been discussed.\cite{1,2,3,4,5,6,7,8,9,10,11,12} It seems that no definitive conclusion has been reached on how $\epsilon_0\partial \v E/\partial t$ should be interpreted.

The idea of treating the displacement current on the same footing as the conduction current stems from the electromagnetism of the nineteenth century. The equation for the vector potential $\v A_C$ in the Coulomb gauge was written at that time as
\begin{equation}
\nabla^2\v A_C=-\mu_0\bigg(\v J+\epsilon_0\frac{\partial\v E}{\partial t}\bigg), \label{1}
\end{equation}
where we have used SI units. The instantaneous solution of Eq.~\eqref{1} is\cite{13}
\begin{equation}
\v A_C(\v x,t)=\frac{\mu_0}{4\pi}\!\int d^3x'\frac{\v J(\v x',t)+\epsilon_0\partial \v E(\v x',t)/\partial t}{R}, \label{2}
\end{equation}
where $R=|\v x'-\v x|$ and the integral is over all space. The displacement current $\epsilon_0\partial \v E/\partial t$ was considered to be equivalent to the conduction current $\v J$. Therefore both $\v J$ and $\epsilon_0\partial \v E/\partial t$ contribute to the magnetic field $\v B$ via the equation $\v B=\gradv \times \v A_C$. The action-at-a-distance ideas prevailing at that time led to the interpretation that both $\v A_C$ and $\v B$ propagate at infinite velocity from their ``true" sources $\v J+\epsilon_0\partial \v E/\partial t$ to the field point.\cite{13}

Similar ideas occur when only fields are considered. From Maxwell's equations we can derive the Poisson equation
\begin{equation}
\nabla^2\v B=-\mu_0\gradv\times(\v J+\epsilon_0\partial\v E/\partial t), \label{Poisson}
\end{equation}
whose solution can be written in the form of the Biot-Savart law generalized to include the displacement current:
\begin{equation}
\v B(\v x,t) = \frac{\mu_0}{4\pi}\!\int\! d^3x'\frac{(\v J(\v x',t) + \epsilon_0\partial \v E(\v x',t)/\partial t)\times{\hat{\v R}}}{R^2}, \label{4}
\end{equation}
where $\hat{\v R} = (\v x -\v x')/R$.

Because Eq.~\eqref{4} is formally correct, two conclusions seem to follow: $\epsilon_0\partial \v E/\partial t$ is a true current like $\v J$, and $\epsilon_0\partial \v E/\partial t$ contributes to the magnetic field. Accordingly, most textbooks interpret the Ampere-Maxwell law,
\begin{equation}
\nabla\times\v B = \mu_0(\v J + \epsilon_0\partial \v E/\partial t),\label{5}
\end{equation}
by stating that ``A changing electric field induces a magnetic field,"\cite{14} which indicates a causal relation between the displacement current and the magnetic field $\v B$.

What is concluded at first sight is not always right. Griffiths and Heald\cite{3} have claimed that Eq.~\eqref{4} is not useful because it seems to be self-referential. To calculate $\v B$ at a point we must know $\epsilon_0\partial \v E/\partial t$; that is, we must know $\v E$ everywhere. But we cannot determine $\v E$ unless we already knows $\v B$ everywhere because of Faraday's induction law
\begin{equation}
\gradv \times \v E=-\partial\v B/\partial t. \label{Faraday}
\end{equation}
Jefimenko\cite{5} and Rosser\cite{13} have pointed out the same circular argument. More importantly, the conclusions can be questioned because Eq.~\eqref{4} involving $\epsilon_0\partial \v E/\partial t$ seems to disagree with two well known properties of electromagnetic phenomena: causality and propagation at the finite speed of light $c$.

The origin of the controversies may be attributed to the fact that most interpretations of $\epsilon_0\partial \v E/\partial t$ do not explicitly consider the retarded solutions of Maxwell's equations. We believe the appropriate interpretation of $\epsilon_0\partial \v E/\partial t$ can be found by taking the time derivative of the retarded electric field. The relation between the displacement current and the conduction current is given by\cite{10}
\begin{equation}
\epsilon_0\frac{\partial \v E}{\partial t}\!=\! -\frac{\v J}{3}+ \!\frac{1}{4\pi}\!\!\int\!\! d^3x'\bigg(\!\frac{3\hat{\v R}(\hat{\v R}\!\cdot\![\v J])\!-\! [\v J]}{R^3}\!
+ \!\frac{3\hat{\v R}(\hat{\v R} \!\cdot\! [\partial \v J/\partial t'])\! -\! [\partial \v J/\!\partial t']}{R^2 c} \!+ \!\frac{\hat{\v R} \!\times \!(\hat{\v R}\! \times\! [\partial^2 \!\v J/\!\partial t'^2])}{Rc^2}\!\bigg), 
\label{7}
\end{equation}
where the brackets $[\;]$ indicate that the enclosed quantity is to be evaluated at the retarded time $t'=t-R/c$. Equation~\eqref{7} will allow us to interpret the displacement current.

In this paper we present a formal interpretation of the displacement current. In Sec.~II we discuss the role played by the displacement current in the context of two traditionally different views: action-at-a-distance and retarded-field-action, and show how Eq.~\eqref{7} clarifies previous interpretations of this current. In Sec.~III we pose the question: Can $\epsilon_0\partial \v E/\partial t$ be interpreted as the ordinary conduction current? We answer this question in the negative because the displacement current 
is a non-localized source while the conduction current is a localized source. In Sec.~IV we pose the question:
Does $\epsilon_0\partial \v E/\partial t$ contributes to the magnetic field? We answer this question in the affirmative. The displacement current effectively contributes to the magnetic field. Similar questions but for the Faraday induction current $(1/\mu_0)\partial \v B/\partial t$ have also been controversial. In Sec.~V we find the relation between $(1/\mu_0)\partial \v B/\partial t$ and $\v J$ [Eq.~\eqref{30}]. 
On the basis of this second relation, we point out that the Faraday induction current is a non-localized source that contributes to the electric field. 
In Sec.~VI we show how the currents $\epsilon_0\partial \v E/\partial t$ and $(1/\mu_0)\partial \v B/\partial t$, both expressed in terms of $\v J$, can be used to implement an instantaneous formulation of Maxwell's equations with local and nonlocal sources. We present our conclusions in Sec.~VII.

\section{The displacement current between action-at-a-distance and retarded-field-action}
It is not difficult to imagine why $\epsilon_0\partial \v E/\partial t$ has been difficult to interpret. There are two relations for the time-dependent magnetic field which seem to be very different. The first one is given by the instantaneous action-at-a-distance expression
\begin{equation}
\v B = \frac{\mu_0}{4\pi}\!\int\! d^3x'\frac{(\v J+\epsilon_0\partial \v E/\partial t)\times{\hat{\v R}}}{R^2}. \label{8}
\end{equation}
The second one is the generalized Biot-Savart law in the retarded form given by Jefimenko\cite{15}
\begin{equation}
\v B =\frac{\mu_0}{4\pi}\!\int\! d^3x'\! \bigg(\frac{[\v J]\times\hat{\v R}}{R^2}+\frac{[\partial\v J/\partial t']\times\hat{\v R}}{R c}\bigg). \label{9}
\end{equation}
Some authors such as Weber and Macomb\cite{1} have argued that the displacement current $\epsilon_0\partial \v E/\partial t$ is like (or equivalent to) $\v J$, contributing to the magnetic field as is shown by Eq.~\eqref{8}. The explicit presence of $\epsilon_0\partial \v E/\partial t$ in Eq.~\eqref{8}
agrees with the common textbook statement that a changing electric field induces a magnetic field. Other authors such as Jefimenko\cite{5} and Rosser\cite{16} have argued that $\epsilon_0\partial \v E/\partial t$ is not like $\v J$ and therefore the former does not contribute to the field $\v B$. Their argument agrees with Eq.~\eqref{9} which states that $\v B$ is produced only by $\v J$ and its time derivative. These authors have reject the idea that a changing electric field produces a magnetic field. The interpretation of what are the true sources for the electric and magnetic fields has been pointed out by Jackson:\cite{9} ``The external charge and current densities are the true sources for the fields." Equation \eqref{9} is consistent with this interpretation. However, Eqs.~\eqref{8} and \eqref{9} are seen to satisfy equivalent equations. Equation~\eqref{8} satisfies Eq.~\eqref {Poisson},\cite{3}
and Eq.~\eqref{9} satisfies the wave equation
\begin{equation}
\Box^2\v B=-\mu_0\gradv\times\v J, \label{10}
\end{equation}
where $\Box^2\equiv\nabla^2-(1/c^2)\partial^2/\partial t^2$. It is easy to show that Eqs.~\eqref{Poisson} and \eqref{10} are the same when Faraday's induction law is used. The equivalence between Eqs.~\eqref{Poisson} and \eqref{10} should not be surprising because Eq.~\eqref{Poisson} involves $\epsilon_0\partial \v E/\partial t$, which is a nonlocal source extending over all space. Therefore 
Eq.~\eqref{Poisson} is really an integro-differential equation. In contrast, Eq.~\eqref{10} is a differential equation involving only the local source $\v J$. Because of the dual description of the field $\v B$ expressed in Eqs.~\eqref{8} and \eqref{9}, the role of the displacement current $\epsilon_0\partial \v E/\partial t$ turns out to be unclear. The dual description of the field $\v B$ in Eqs.~\eqref{8} and \eqref{9} is also given for the field $\v E$ as we will see in Sec.~V.

It is interesting that a dual formulation (instantaneous and retarded) was already known in the nineteenth century for potentials. The Coulomb-gauge vector potential that generates the field $\v B$ in Eq.~\eqref{8} is given by $\v A_C=\mu_0\!\int\! d^3x'(\v J+\epsilon_0\partial \v E/\partial t)/(4\pi R)$, and the Lorenz-gauge vector potential $\v A_L$ that generates the field $\v B$ in Eq.~\eqref{9} is given by $\v A_L=\mu_0\!\int\! d^3x'[\v J]/(4\pi R)$. Henri Poincare in 1894 pointed out\cite{17} ``In calculating [$\v A_C$] Maxwell takes into account the currents of
conduction and those of displacement; and he supposes that the attraction takes place
according to Newton's law, i.e. instantaneously. But in calculating [$\v A_L$] on the contrary we take account only of conduction currents and we suppose the attraction is propagated with the velocity of light \ldots It is a matter of indifference
whether we make this hypothesis [of a propagation in time] and consider only the induction due to conduction currents, or whether like Maxwell, we retain the old law of [instantaneous] induction and consider both conduction and the displacement currents."

Equation~\eqref{7} allows us to understand the displacement current $\epsilon_0\partial \v E/\partial t$ from a different perspective. In Appendix~A we derive Eq.~\eqref{7} from the retarded electric field expressed in terms of Lorenz-gauge potentials. The direct interpretation of Eq.~\eqref{7} is simple. The first term of $\epsilon_0\partial \v E/\partial t$ in Eq.~\eqref{7} represents a local contribution that depends on the present value of $\v J$. The second term $\epsilon_0\partial \v E/\partial t$ is a nonlocal contribution that extends over all space and depends on the retarded values of $\v J$. The local contribution to $\epsilon_0\partial \v E/\partial t$ is instantaneously produced by $\v J$ and the nonlocal contribution to $\epsilon_0 \partial \v E/\partial t$ is causally produced by $\v J$. We conclude that the displacement current is produced by the present and retarded values of the conduction current $\v J$. The question then arises: Is $\epsilon_0 \partial \v E/\partial t$ a real source? According to Eq.~\eqref{7} the displacement current is constructed from the real external current density. Therefore Maxwell was not wrong when he called $\v J+\epsilon_0\partial \v E/\partial t$ ``the true electric current." Equation~\eqref{7} shows that the displacement current is as real as is the conduction current because the former is ultimately constructed from the latter. In particular, if $\v J=0$, then $\epsilon_0\partial \v E/\partial t=0$.
The essential difference between the currents is that the displacement current involves a nonlocal part, and the conduction current is a localized source. If Maxwell had known Eq.~\eqref{7}, he could have expressed the ``true" electric current $\v J+ \epsilon_0\partial \v E/\partial t$ entirely in terms of $\v J$, that is, $\v J+ \epsilon_0\partial \v E/\partial t=2\v J/3 + \int d^3 x'f(\v J)$.

The new relation in Eq.~\eqref{7} improves our understanding of the displacement current. For example, the self-referential character of Eq.~\eqref{4}, discussed in Refs.~\onlinecite{3, 5,13} is removed by Eq.~\eqref{7}. To calculate $\v B$ using Eq.~\eqref{4} we must know $\epsilon_0 \partial \v E/\partial t$, which can be calculated from $\v J$ using Eq.~\eqref{7} with no explicit reference to $\v B$.

Equation~\eqref{7} allows us to clarify various points of view regarding the displacement current. For example, Arthur\cite{12} has claimed that ``The displacement current is not like a real current but, in effect, an artifact of Maxwell's equations that may be treated as a sort of current \ldots" Equation~\eqref{7} shows that the displacement current is not an artifact; we can assign a meaning to the current $\epsilon_0 \partial \v E/\partial t$ in terms of the observed conduction current $\v J$. According to Giuliani,\cite{18} a theoretical term has
a physical meaning if it cannot be withdrawn without reducing the predictive
power of the theory. In our context we cannot eliminate the displacement current without reducing the predictive power of Maxwell's equations. Given this criterion, the displacement current has a physical meaning which is made transparent through Eq.~\eqref{7}.

Note that the right-hand side of Eq.~\eqref{7} displays the characteristic form of the electric field produced by a time-dependent polarization density $\v P$. In Ref.~\onlinecite{10} we have shown that the substitution $\v J=\partial \v P/\partial t$ in Eq.~\eqref{7} implies the electric field produced by the polarization vector $\v P$. Therefore 
Eq.~\eqref{7} with $\v J=\partial \v P/\partial t$ can be used to obtain, for example, the Hertz fields of an electric dipole.

Equation~\eqref{8} can also be obtained by applying the standard Helmholtz theorem extended to include time dependence. According to this instantaneous formulation of the theorem, an instantaneous vector field $\v F(\v x,t)$ which vanishes at infinity is determined by specifying its divergence and curl. An expression that illustrates this theorem is given by
\begin{equation}
\v F =-\gradv\!\int\! d^3x'\frac{\gradv'\cdot\v F}{4\pi R}+\gradv\times\!\int\! d^3x'\frac{\gradv'\times \v F}{4\pi R}. \label{11}
\end{equation}
If we apply this theorem to the magnetic field $\v B$, that is, $\v F=\v B$, and use Maxwell's equations
\begin{equation}
\gradv\cdot\v B=0\ \mbox{and}\ 
\gradv\times \v B=\mu_0(\v J+\epsilon_0\partial \v E/\partial t), \label{maxwell}
\end{equation}
we obtain
\begin{equation}
\v B =\frac{\mu_0}{4\pi}\gradv\times\!\int\! d^3x'\frac{\v J}{R}+\frac{\mu_0}{4\pi}\gradv\times\!\int\! d^3x'\frac{\epsilon_0\partial \v E/\partial t}{R}. \label{13}
\end{equation}
After an integration by parts, Eq.~\eqref{13} reduces to Eq.~\eqref{8}.

The Helmholtz theorem extended to include time dependence in a retarded form allows us to obtain Eq.~\eqref{9}. According to this formulation of the theorem,\cite{19,20} a retarded vector field $\v F(\v x,t)$ vanishing at infinity is completely determined by specifying its divergence, curl, and time derivative. An expression that illustrates this theorem is given by\cite{19}
\begin{equation}
\v F =-\gradv\!\int\! d^3x'\frac{[\gradv'\cdot\v F]}{4\pi R}+\gradv\times\!\int\! d^3x'\frac{[\gradv'\times \v F]}{4\pi R}+\frac{1}{c^2}\frac{\partial}{\partial t}
\!\int\! d^3x'\frac{[\partial \v F/\partial t']}{4\pi R}. \label{14}
\end{equation}
If we let $\v F=\v B$ and use Eqs.~\eqref{Faraday} and \eqref{maxwell}, we obtain
\begin{equation}
\v B =\gradv\times\!\int\! d^3x'\frac{[\mu_0\v J+(1/c^2)\partial\v E/\partial t']}{4\pi R}-\frac{1}{c^2}\frac{\partial}{\partial t}
\!\int\! d^3x'\frac{[ \gradv'\times \v E]}{4\pi R}. \label{15}
\end{equation}
An integration by parts and the property $\partial [\;]/\partial t = [\partial/\partial t']$ yields
\begin{equation}
\gradv\times\!\int\! d^3x'\frac{[\partial\v E/\partial t']}{4\pi c^2 R}= \frac{1}{c^2}\frac{\partial}{\partial t}
\!\int\! d^3x'\frac{[ \gradv'\times \v E]}{4\pi R}. \label{16}
\end{equation}
We use Eqs.~\eqref{15} and \eqref{16} and obtain
\begin{equation}
\v B =\frac{\mu_0}{4\pi}\gradv\times\!\int\! d^3x'\frac{[\v J]}{R}. \label{17}
\end{equation}
After an integration by parts, Eq.~\eqref{17} reduces to Eq.~\eqref{9}.\cite{3} We proceed now to answer the questions posed at the end of the introduction in Sec.~I.

\section{Can $\epsilon_0\partial \v E/\partial t$ be interpreted as the ordinary conduction current?}

Despite the fact that the
displacement current $\epsilon_0\partial \v E/\partial t$ is constructed with the conduction current $\v J$ as is shown in Eq.~\eqref{7}, it cannot be interpreted in the same way as the current $\v J$ because $\v J$ is a local source and the current $\epsilon_0\partial \v E/\partial t$ contains a nonlocal contribution.

We define ordinary sources as those satisfying at least two requirements: (a) they should not originate self-referential expressions for their associated fields (a self-referential expression for a field $\v F$ involves a source $\bff$ which is ultimately determined by $\v F$ itself), and (b) they should be physically realizable as is the conduction current. This condition means that ordinary sources are localized in finite regions of space. Artificial configurations such as an infinite wire or an infinite plane are nonlocal sources that can be considered as approximations to ordinary sources. The displacement current satisfies requirement (a) but not requirement (b), and therefore this current cannot be interpreted as an ordinary source.

\section{Does $\epsilon_0\partial \v E/\partial t$ produce (or contribute to) a magnetic field?}

When $\epsilon_0\partial \v E/\partial t$ is expressed in terms of $\v J$ [as shown in Eq.~\eqref{7}] and substituted in Eq.~\eqref{4}, the current $\epsilon_0\partial \v E/\partial t$ contributes to the magnetic field. To see this result, first consider the potential $\v A_C$. If we insert Eq.~\eqref{7} into Eq.~\eqref{1}, we find a novel form for $\v A_C$ entirely in terms of the current $\v J$:
\begin{align}
\v A_C(\v x,t)&= \frac{\mu_0}{6\pi}\!\int\! d^3x'\frac{\v J(\v x',t)}{R}
+\frac{\mu_0}{4\pi}\!\int\!\! d^3x''\!\frac{1}{R'}\bigg\{\!\frac{1}{4\pi}\!\!\int\!\! d^3x'\!\bigg(\!\frac{3\hat{\v R}''(\hat{\v R}'' \cdot [\v J]^*) - [\v J]^*}{R''^3}\nonumber\\
&{} \quad + \frac{3\hat{\v R}''(\hat{\v R}''\! \cdot \![\partial \v J/\partial t']^*) - [\partial \v J/\partial t']^*}{R''^2 c}
+ \frac{\hat{\v R}'' \times (\hat{\v R}'' \times [\partial^2 \v J/\partial t'^2]^*)}{R''c^2}\bigg)\!\bigg\}, \label{18}
\end{align}
where $R'=|\v x-\v x''|$, $R''=|\v x''-\v x'|$, $\hat{\v R}''=(\v x''-\v x')/R''$, and the brackets $[~]^*$ indicate that the enclosed quantity is to be evaluated at the retarded time $t'=t-R''/c$. The first term in Eq.~\eqref{18} is an instantaneous contribution. The integrand of $\!\int d^3x'$ in the second term of Eq.~\eqref{18} involves retardation. After performing the specified integration in the braces $\{~\}$, the resulting quantity $\v Y (\v x'', t)$
is a function of $\v x''$ and the present time $t$. In terms of $\v Y (\v x'', t)$ the second term is $(\mu_0/(4\pi)\!\int\! d^3x''\v Y (\v x'', t)/R'$. Because $\v x''$ is a dummy variable, it can be changed to $\v x'$, and thus we can write the second term as an effective instantaneous contribution $(\mu_0/(4\pi))\!\int\! d^3x'\v Y (\v x', t)/R$. We then obtain an instantaneous expression for $\v A_C$:
\begin{equation}
\v A_C(\v x,t)=\frac{\mu_0}{6\pi}\!\int\! d^3x'\frac{\v J(\v x',t) + 3\v Y(\v x',t)/2}{R}. \label{19}
\end{equation}
Similarly, we insert Eq.~\eqref{7} into Eq.~\eqref{2} to obtain a novel expression for the magnetic field:
\begin{align}
\v B(\v x,t) &= \frac{\mu_0}{6\pi}\!\int\! d^3x' \frac{\v J(\v x',t)\times \hat{\v R}}{R^2}
- \frac{\mu_0}{4\pi}\!\int\! d^3x''\!\frac{\hat{\v R}'}{R'^2}\times\!\bigg\{\!\frac{1}{4\pi}\!\int\!\! d^3x'\!\bigg(\!\frac{3\hat{\v R}''(\hat{\v R}'' \cdot [\v J]^*) - [\v J]^*}{R''^3}\nonumber\\
&{}\quad
+ \frac{3\hat{\v R}''(\hat{\v R}'' \cdot [\partial \v J/\partial t']^*) - [\partial \v J/\partial t']^*}{R''^2 c}
+ \frac{\hat{\v R}'' \times (\hat{\v R}'' \times [\partial^2 \v J/\partial t'^2]^*)}{R''c^2}\!\bigg\}. \label{20}
\end{align}
The first term in Eq.~\eqref{20} is an instantaneous contribution. Following the previous analysis for $\v A_C$, the second term in Eq.~\eqref{20} also describes an effective instantaneous contribution: $(\mu_0/4\pi)\!\int\! d^3x' \v Y (\v x', t)\times\hat{\v R}/R$. Therefore Eq.~\eqref{20} represents an instantaneous expression for the magnetic field which can be written compactly as
\begin{equation}
\v B(\v x,t) = \frac{\mu_0}{6\pi}\!\int\!\! d^3x'\frac{(\v J(\v x',t) + 3\v Y(\v x',t)/2)\times\hat{\v R}}{R^2}. \label{21}
\end{equation}
Equations~\eqref{18} and \eqref{20} [or equivalently Eqs.~\eqref{19} and \eqref{21}] are related by $\v B=\gradv \times \v A_C$.

\section{Does $(1/\!\mu_0)\partial\v B/\!\partial t$ produce (or contribute to) an electric field?}

The question of whether $(1/\mu_0)\partial\v B/\partial t$ produces an electric field arises from a direct interpretation of Faraday's law when it is expressed as in Eq.~\eqref{Faraday}.
Some textbooks\cite{14} write that according to this law ``A changing magnetic field induces an electric field." This statement indicates a causal relation between $\partial \v B/\partial t$ and $\v E$. From the Gauss and Faraday laws we can derive the equation
\begin{equation}
\nabla^2\v E=\gradv\rho/\epsilon_0+\mu_0\gradv\times\bigg(\frac{1}{\mu_0} \frac{\partial \v B}{\partial t} \bigg), \label{22}
\end{equation}
where $(1/\mu_0)\partial \v B/\partial t$ can be considered to be a current density which we call the ``induction current." Note that the induction and displacement currents do not have the same units. That is, the units of the induction current are m/s times the units of the displacement current. The instantaneous solution of Eq.~\eqref{22} can be written in the form of Coulomb's law generalized to include the induction current:
\begin{equation}
\v E(\v x,t) = \frac{1}{4\pi\epsilon_0}\!\int\!\!d^3\!x'\!\bigg(\!\frac{\rho(\v x',t)\hat{\v R}}{R^2}
-\frac{\!((1/\mu_0)\partial \v B(\v x',t)/\partial t)\times\hat{\v R}}{R^2c^2} \!\bigg). \label{23}
\end{equation}
We might conclude that $(1/\mu_0)\partial \v B/\partial t$ plays a role similar to that of the charge density $\rho$, 
that is, $(1/\mu_0)\partial \v B/\partial t$ is a true current which contributes to the electric field. But this conclusion is suspect because the electric field can also be written as the time-dependent generalization of Coulomb's law in the retarded form given by Jefimenko:\cite{15}
\begin{equation}
\v E =\frac{1}{4\pi\epsilon_0}\!\int d^3x'\bigg(\frac{\hat{\v R}}{R^2}[\rho]+\frac{\hat{\v R}}{Rc}\left[\frac{\partial \rho}{\partial t}\right]
-\frac{1}{Rc^2}\left[\frac{\partial \v J}{\partial t}\right]\bigg).\label{24}
\end{equation}
According to Eq.~\eqref{24}, the charge and current densities are the true sources for the electric field. The absence of $(1/\mu_0)\partial \v B/\partial t$ in Eq.~\eqref{24} supports the claims by Jefimenko\cite{5} and Rosser\cite{21} who have rejected the idea that a changing magnetic field induces an electric field. However, Eqs.~\eqref{23} and \eqref{24} satisfy relations that are formally equivalent. On one hand, Eq.~\eqref{23} satisfies the Poisson equation in Eq.~\eqref{22}. On the other hand, Eq.~\eqref{24} satisfies the wave equation
\begin{equation}
\Box^2\v E=\gradv\rho/\epsilon_0 + \mu_0\partial\v J/\partial t. \label{25}
\end{equation}
It is easy to see that Eqs.~\eqref{22} and \eqref{25} are the same when the Ampere-Maxwell law is used. The equivalence of Eqs.~\eqref{22} and \eqref{25} is not unexpected because Eq.~\eqref{22} involves $\mu_0\partial \v B/\partial t$, which is a nonlocal source extending over all space. Therefore  
Eq.~\eqref{22} is an integro-differential equation. In contrast, Eq.~\eqref{25} is a differential equation with the local sources $\rho$ and $\v J$. Note that Eq.~\eqref{23} can also be obtained by applying the instantaneous formulation of Helmholtz's theorem in Eq.~\eqref{11}. If we apply Eq.~\eqref{11} to the electric field $\v E$ and use Maxwell's equations $\gradv\cdot \v E=\rho/\epsilon_0$ and $\gradv\times\v E=-\partial \v B/\partial t$, we obtain
\begin{equation}
\v E=-\frac{1}{4\pi\epsilon_0}\gradv\!\int d^3x'\frac{\rho}{R}-\frac{\mu_0}{4\pi}\gradv \times \!\int\!d^3x'\frac{(1/\mu_0)\partial \v B/\partial t}{R}. \label{26}
\end{equation}
After an integration by parts, Eq.~\eqref{26} yields Eq.~\eqref{23}.

Equation~\eqref{24} can be derived by applying the retarded formulation of Helmholtz's theorem in Eq.~\eqref{14}. If we apply Eq.~\eqref{14} to the magnetic field $\v E$ and use Maxwell's equations $\gradv\cdot\v E=\rho/\epsilon_0$, $\gradv\times \v E=-\partial \v B/\partial t$, and $\partial \v E/\partial t=c^2\gradv\times \v B-\mu_0c^2\v J$, we obtain
\begin{equation}
\v E =-\gradv\!\int\! d^3x'\frac{[\rho]}{4\pi\epsilon_0 R}-\gradv\times\!\int\! d^3x'\frac{[\partial\v B/\partial t']}{4\pi R}+\frac{\partial}{\partial t}
\!\int\! d^3x'\frac{[ \gradv'\times \v B-\mu_0\v J]}{4\pi R}. \label{27}
\end{equation}
An integration by parts and the relation $\partial [~]/\partial t = [\partial/\partial t']$ yields the result
\begin{equation}
-\gradv\times\!\int\! d^3x'\frac{[\partial\v B/\partial t']}{4\pi R}= -\frac{\partial}{\partial t}
\!\int\! d^3x'\frac{[ \gradv'\times \v B]}{4\pi R}. \label{28}
\end{equation}
If we use Eqs.~\eqref{27} and \eqref{28}, we obtain
\begin{equation}
\v E =-\frac{1}{4\pi\epsilon_0}\gradv\!\int\! d^3x'\frac{[\rho]}{ R}-\frac{\mu_0}{4\pi}\frac{\partial}{\partial t}\int\! d^3x'\frac{[\v J]}{R}. \label{29}
\end{equation}
After an integration by parts, Eq.~\eqref{29} reduces to Eq.~\eqref{24}.\cite{3}

There is also a relation between the currents $(1/\mu_0)\partial \v B/\partial t$ and $\v J$. By taking the time derivative of Eq.~\eqref{9} and using $\partial [~]/\partial t=[\partial/\partial t']$, we obtain an expression for the induction current entirely in terms of $\v J$:
\begin{equation}
\frac{1}{\mu_0}\frac{\partial \v B}{\partial t} = \frac{1}{4\pi}\!\int\!\! d^3x'\!\bigg(\frac{[\partial \v J/\partial t']\times\hat{\v R} }{R^2}
+\!\frac{[\partial^2 \v J/\partial t'^2]\times\hat{\v R} }{Rc}\!\bigg). \label{30}
\end{equation}
Equation~\eqref{30} states that the induction current $(1/\mu_0)\partial \v B/\partial t$ is causally produced by the conduction current $\v J$. We can say that the current $(1/\mu_0)\partial \v B/\partial t$ is a true source because it is constructed from the true source $\v J$. If $\v J=0$, then $(1/\mu_0)\partial \v B/\partial t=0$. However, $(1/\mu_0)\partial \v B/\partial t$ is a nonlocal source and cannot be interpreted as the local conduction current $\v J$.

To answer the question about whether $(1/\mu_0)\partial \v B/\partial t$ produces or contributes to an electric field, we substitute Eq.~\eqref{30} into Eq.~\eqref{23} to obtain a novel expression for the electric field:
\begin{align}
\v E(\v x,t) & = \frac{1}{4\pi\epsilon_0}\!\int\!d^3x'\frac{\rho(\v x',t)\hat{\v R}}{R^2}
+\frac{1}{4\pi \epsilon_0 c^2}\!\int\!\! d^3x''\!\frac{\hat{\v R'}}{R'^2}\times\bigg\{\!\frac{1}{4\pi}\!\int\!d^3x'\!\bigg(\frac{[\partial \v J/\partial t']^*\times\hat{\v R''} }{R''^2}\nonumber\\
&{}\quad +\frac{[\partial^2 \v J/\partial t'^2]^*\times\hat{\v R''} }{R''c}\!\bigg)\!\bigg\}. \label{31}
\end{align}
The first term in Eq.~\eqref{31} is the instantaneous Coulomb field. The second term involves retardation. After performing the integration
in the braces $\{~\}$, the resulting quantity $\v Z (\v x'', t)$ is a function of $\v x''$ and the present time $t$. In terms of $\v Z (\v x'', t)$ the second term in Eq.~\eqref{31} can be written as $-(1/(4\pi c^2))\!\int\! d^3x''\v Z (\v x'', t)\times \hat{\v R}'/R'^2$.
The variable $\v x''$ is a dummy variable and can be changed to $\v x'$, so that the second term can be expressed as $-(1/(4\pi c^2))\!\int\! d^3x'\v Z (\v x', t)\times\hat{\v R}/R^2$, indicating that it is an effective instantaneous contribution.
We conclude that Eq.~\eqref{31} is an instantaneous expression for the electric field and can be written compactly as
\begin{equation}
\v E(\v x,t) = \frac{1}{4\pi\epsilon_0}\!\int\!\! d^3x'\frac{\rho(\v x',t)\hat{\v R} - (1/c^2)\v Z(\v x',t)\times\hat{\v R}}{R^2}. \label{32}
\end{equation}

We now answer the question posed in the title of this section. The answer is yes if the explicit form of the induction current $(1/\mu_0)\partial \v B/\partial t$ given by Eq.~\eqref{30} is inserted into Eq.~\eqref{23}. Then this current clearly contributes to the electric field.

\section{Instantaneous formulation of Maxwell's equations with local and nonlocal sources}

Maxwell's equations imply wave equations [Eq.~\eqref{10} and Eq.~\eqref{25}] with local sources $\rho$ and $\v J$. The retarded solutions of 
Eq.~\eqref{10} and Eq.~\eqref{25} are given by Eq.~\eqref{9} and Eq.~\eqref{24} which display causality and propagation at the speed of light $c$. The retarded formulation of Maxwell's equations is well known. But Maxwell's equations also imply Poisson equations [Eq.~\eqref{Poisson} and Eq.~\eqref{22}]
with local sources $\rho$ and $\v J$ as well as nonlocal sources $\epsilon_0 \partial\v E/\partial t$ and $(1/\mu_0)\partial\v B/\partial t$. The instantaneous solutions of Eq.~\eqref{Poisson} and Eq.~\eqref{22} are given by Eq.~\eqref{4} and Eq.~\eqref{23} which display acausality and infinite propagation. However, the ``instantaneous" formulation of Maxwell's equations is not well known. It is therefore appropriate to discuss the instantaneous formulation here.

For convenience we write the displacement and induction currents as
\begin{align}
\epsilon_0\frac{\partial \v E}{\partial t}&= -\frac{\v J}{3} +\v Y, \label{33}\\
\frac{1}{\mu_0}\frac{\partial \v B}{\partial t} &= \v Z, \label{34}
\end{align}
where the time-dependent vectors $\v Y(\v x,t)$ and $\v Z(\v x,t)$ are not specified. By using Eqs.~\eqref{33} and \eqref{34} and Maxwell's equations, we can derive the relations
\begin{align}
\gradv\cdot\v Y&=-\frac{2}{3}\nabla\cdot\v J, \label{35}\\
\gradv\cdot\v Z&= 0, \label{36} \\
\gradv\times \v Y&=\frac{1}{3} \nabla\times\v J-\epsilon_0\mu_0\frac{\partial \v Z}{\partial t}, \label{37}\\
\gradv\times \v Z
&=\frac{2}{3}\frac{\partial \v J}{\partial t}+\frac{\partial \v Y}{\partial t}, \label{38}
\end{align}
which imply the wave equations:
\begin{align}
\Box^2\bigg(\v Y +\frac{2}{3}\v J\bigg)&= -\gradv\times(\gradv\times\v J), \label{39} \\
\Box^2\v Z &= -\frac{\partial}{\partial t}\gradv \times \v J. \label{40}
\end{align}
The retarded solutions of Eqs.~\eqref{39} and \eqref{40} are given by
\begin{align}
\v Y &= -\frac{2}{3}\:\v J+ \frac{1}{4\pi}\!\int\! d^3x'\frac{[\gradv'\times(\gradv'\times\v J)]}{R}, \label{41} \\
\v Z&=\frac{1}{4\pi}\!\int\! d^3x'\frac{[\gradv' \times \partial \v J/\partial t']}{R}. \label{42}
\end{align}
Therefore the vectors $\v Y$ and $\v Z$ can be calculated from the current density $\v J$.

An alternative form of Eqs.~\eqref{41} and \eqref{42} is given by
\begin{align}
\v Y&= \frac{1}{4\pi}\!\int \!d^3x' \bigg(\!\frac{3\hat{\v R}(\hat{\v R} \cdot [\v J]) - [\v J]}{R^3} + \frac{3\hat{\v R}(\hat{\v R} \cdot [\partial \v J/\partial t']) - [\partial \v J/\partial t']}{R^2 c}
+ \frac{\hat{\v R} \times (\hat{\v R} \times [\partial^2 \v J/\partial t'^2])}{Rc^2}\bigg),\label{43} \\
\v Z&= \frac{1}{4\pi}\!\int\! d^3x'\!\bigg(\!\frac{[\partial \v J/\partial t']\times\hat{\v R} }{R^2} + \frac{[\partial^2 \v J/\partial t'^2]\times\hat{\v R} }{Rc}\bigg). \label{44}
\end{align}
Equation \eqref{44} has been derived in Eq.~\eqref{30} because $\v Z=(1/\mu_0)\partial\v B/\partial t$. Equation~\eqref{43} is demonstrated in Appendix~A.

By using Eqs.~\eqref{33} and \eqref{34}, Maxwell's equations become a system of equations with local sources $\rho$ and $\v J$ as well as nonlocal sources $\v Y$ and $\v Z$:
\begin{align}
\gradv\cdot\v E&=\rho/\epsilon_0, \label{45}\\
\gradv\cdot\v B&= 0,\label{46}\\
\gradv\times \v E&=-\mu_0\v Z, \label{47}\\
\gradv\times \v B&=\frac{2\mu_0}{3}\v J+\mu_0\v Y.\label{48}
\end{align}
Because $\rho$ and $\v J$ are specified sources and $\v Y$ and $\v Z$ can be determined from $\v J$, it follows that all sources on the right-hand side of Eqs.~\eqref{45}--\eqref{48} are known functions of space and time. Note that the electric and magnetic fields are explicitly decoupled in Eqs.~\eqref{45}--\eqref{48}.

Maxwell's equations in the form given by Eqs.~\eqref{45}--\eqref{48} represent an instantaneous action-at-a-distance theory. From these equations we obtain the Poisson equations:
\begin{align}
\nabla^2\v E&=\gradv\rho/\epsilon_0+\mu_0\gradv\times\v Z,\label{49}\\
\nabla^2\v B&=-\frac{2\mu_0}{3}\gradv\times(\v J + 3\v Y/2),\label{50}
\end{align}
whose instantaneous solutions are given by Eqs.~\eqref{32} and \eqref{21}, respectively.

The instantaneous formulation of Maxwell's equations in Eqs.~\eqref{45}--\eqref{48} does not imply that causality and propagation at speed of light $c$ have disappeared from these equations. The sources $\v Y$ and $\v Z$ are determined from the retarded values of the current $\v J$. The quantities $\v Y-\v J/3$ and $\v Z$ satisfy wave equations propagating at speed of light $c$. We have here a formal dilemma. If, as usual, only the local sources $\rho$ and $\v J$ are used, then Maxwell's equations are a set of differential equations yielding retarded solutions. But if the nonlocal sources $\v Y$ and $\v Z$ are considered in addition to the local sources $\rho$ and $\v J$, then Maxwell's equations are a set of integro-differential equations yielding instantaneous solutions. In other words, the price for having an instantaneous formulation of Maxwell's equations consists in introducing the nonlocal sources $\v Y$ and $\v Z$ which are causally produced by the local current $\v J$. As noted, Poincare\cite{17} was probably the first to point out the dual formulation (instantaneous and retarded) of Maxwell's equations. The instantaneous formulation was forgotten because of the emergence of special relativity.\cite{22}

\section{Conclusion}
The question on whether the displacement current $\epsilon_0\partial \v E/\partial t$
produces a magnetic field and represents a true current like the conduction current $\v J$ can be answered by considering Eq.~\eqref{7}, which expresses $\epsilon_0\partial \v E/\partial t$ in terms of $\v J$. A similar controversy on whether the Faraday induction current $(1/\mu_0)\partial \v B/\partial t$ produces an electric field and represents a true source can be clarified by means of Eq.~\eqref{30}, which expresses $(1/\mu_0)\partial \v B/\partial t$ in terms of $\v J$. We have shown that the current $\epsilon_0\partial \v E/\partial t$ in Eq.~\eqref{7} contributes to the field $\v B$ and that the current $(1/\mu_0)\partial \v B/\partial t$ in Eq.~\eqref{30} contributes to the field $\v E$. However, both $\epsilon_0\partial \v E/\partial t$ and $(1/\mu_0)\partial \v B/\partial t$ cannot be considered as ordinary sources like $\v J$ because they are nonlocal. The use of Eqs.~\eqref{7} and \eqref{30} allowed us to implement an instantaneous formulation of Maxwell's equations which is given in Eqs.~\eqref{45}--\eqref{48}. A dual formulation of Maxwell's equations is then feasible: If $\epsilon_0\partial \v E/\partial t$ and $(1/\mu_0)\partial \v B/\partial t$ are used as nonlocal sources, then Maxwell's equations become a set of integro-differential equations representing an instantaneous action-at-a-distance theory. If only the local sources $\rho$ and $\v J$ are used, then Maxwell's equations are a set of differential equations representing a retarded field theory. Both formulations are formally equivalent.

\appendix
\section{Derivation of Eq.~\eqref{7}}
When the Lorenz-gauge potentials
\begin{equation}
\Phi_L=\frac{1}{4\pi\epsilon_0}\int d^3 x'\frac{[\rho]}{R}\ \mbox{and}\ \v A_L=\frac{\mu_0}{4\pi}\int d^3 x'\frac{[\v J]}{R},\label{a1}
\end{equation}
are inserted in the electric field $\v E=-\gradv\Phi_L-\partial\v A_L/\partial t$ we obtain Eq.~\eqref{29}.
The time derivative of Eq.~\eqref{29}, $\partial [~]/\partial t=[\partial/\partial t'],\mu_0\epsilon_0 = 1/c^2$ and the continuity equation yield
\begin{equation}
\epsilon_0\frac{\partial \v E}{\partial t} = \frac{1}{4\pi}\gradv\!\int\! d^3x'\frac{[\gradv'\cdot\v J]}{R}-\frac{1}{4\pi c^2}\!\int\! d^3x'\frac{[\partial^2\v J/\partial t'^2]}{R}.\label{a2}
\end{equation}
After an integration by parts, we obtain
\begin{equation}
\int\! d^3x'\frac{[\gradv'\cdot\v J]}{R}=\int\! d^3x'\gradv\cdot\bigg(\frac{[\v J]}{R}\bigg)+\oint\!d\v S'\cdot\frac{[\v J]}{R}.\label{a3}
\end{equation}
The surface integral vanishes because $\v J$ is assumed to be a localized source. We use Eqs.~\eqref{a2} and \eqref{a3} to write
\begin{equation}
\epsilon_0\frac{\partial \v E}{\partial t} = \frac{1}{4\pi}\!\int\!\! d^3x'\gradv\bigg(\!\gradv\cdot\bigg(\frac{[\v J]}{R}\bigg)\!\bigg) - \frac{1}{4\pi c^2}\!\int\! d^3x'\frac{[\partial^2\v J/\partial t'^2]}{R}.\label{a4}
\end{equation}

We next demonstrate the identity:
\begin{align}
\gradv\bigg(\!\gradv\cdot\bigg(\frac{[\v J]}{R}\bigg)\!\bigg)&=\bigg(\frac{3\hat{\v R}(\hat{\v R}\cdot [\v J]) - [\v J]}{R^3}
+ \frac{3\hat{\v R}(\hat{\v R} \cdot [\partial \v J/\partial t']) - [\partial \v J/\partial t']}{R^2 c}
+ \frac{\hat{\v R}(\hat{\v R}\cdot[\partial^2 \v J/\partial t'^2])}{Rc^2} \bigg)\nonumber\\
&{}\quad -\frac{4\pi}{3}\v [\v J]\delta(\v x-\v x'),\label{a5}
\end{align}
where $\delta(\v x-\v x')$ is the Dirac delta function. Consider the left-hand side of Eq.~\eqref{a5}:
\begin{align}
\gradv\bigg(\!\gradv\cdot\bigg(\frac{[\v J]}{R}\bigg)\!\bigg)=\frac{\gradv(\gradv\cdot[\v J])}{R}+(\gradv\cdot[\v J])\gradv\bigg(\!\frac{1}{R}\!\bigg) + \gradv\bigg(\![\v J]\cdot\gradv\bigg(\!\frac{1}{R}\!\bigg)\!\bigg).\label{a6}
\end{align}
The first term in Eq.~\eqref{a6} can be written as
\begin{equation}
\frac{\gradv(\gradv\cdot[\v J])}{R}= \frac{\hat{\v R}(\hat{\v R}\cdot[\partial\v J/\partial t'])-[\partial\v J/\partial t']}{R^2c}
+\frac{\hat{\v R}(\hat{\v R}\cdot[\partial^2\v J/\partial t'^2])}{Rc^2},\label{a7}
\end{equation}
where we have used the relations
\begin{align}
\gradv\cdot[\v J] &= -\frac{\hat{\v R}\cdot[\partial \v J/\partial t']}{c},\label{a8}\\
\gradv\times[\partial\v J/\partial t'] &= -\frac{\hat{\v R}\times[\partial^2 \v J/\partial t'^2]}{c},\label{a9}\\
(\hat{\v R}\cdot\gradv)[\partial \v J/\partial t']&=-\frac{[\partial^2 \v J/\partial t'^2]}{c},\label{a10}\\
([\partial\v J/\partial t']\cdot\gradv)\hat{\v R}&=-\frac{\hat{\v R}(\hat{\v R}\cdot[\partial\v J/\partial t'])-[\partial\v J/\partial t']}{R}.\label{a11}
\end{align}
The second term in Eq.~\eqref{a6} can be expressed as
\begin{equation}
(\gradv\cdot[\v J])\gradv\bigg(\!\frac{1}{R}\!\bigg)=\frac{\hat{\v R}(\hat{\v R}\cdot[\partial\v J/\partial t'])}{R^2c},\label{a12}
\end{equation}
where we have used Eq.~\eqref{a8}. The third term in Eq.~\eqref{a6} can be expressed as
\begin{equation}
\gradv\bigg(\![\v J]\!\cdot\!\gradv\bigg(\!\frac{1}{R}\!\bigg)\!\bigg)=\frac{3\hat{\v R}(\hat{\v R}\cdot [\v J]) - [\v J]}{R^3} + \frac{\hat{\v R}(\hat{\v R} \cdot [\partial \v J/\partial t'])}{R^2 c} -\frac{4\pi}{3}\v [\v J]\delta(\v x-\v x'),\label{a13}
\end{equation}
where we have used the relations
\begin{align}
\gradv\times[\v J] &= -\frac{\hat{\v R}\times[\partial\v J/\partial t']}{c},\label{a14}\\
([\v J]\cdot\gradv)\gradv\bigg(\!\frac{1}{R}\!\bigg)&=\frac{3\hat{\v R}(\hat{\v R}\cdot [\v J]) - [\v J]}{R^3} -\frac{4\pi}{3}\v [\v J]\delta(\v x-\v x'),\label{a15}\\
\bigg(\!\gradv\bigg(\!\frac{1}{R}\!\bigg)\!\cdot\!\gradv\!\bigg)[\v J]&=\frac{[\partial^2 \v J/\partial t'^2])}{R^2 c}.\label{a16}
\end{align}
To obtain Eq.~\eqref{a15} we have applied the identity
\begin{equation}
\frac{\partial}{\partial x_i\partial x_j}\bigg(\!\frac{1}{R}\!\bigg)=\frac{3\hat{R_i}\hat{R_j}-\delta_{ij}}{R^3} -\frac{4\pi}{3}\delta_{ij}\delta(\v x-\v x'),\label{a17}
\end{equation}
where $\hat{R_i}=R_i/R=(x_i-x'_i)/R$ and $\delta_{ij}$ is the Kronecker delta symbol. If Eqs.~\eqref{a7}, \eqref{a12}, and \eqref{a13} are substituted in Eq.~\eqref{a6}, we obtain the identity in Eq.~\eqref{a5}.

The volume integration of Eq.~\eqref{a5} yields
\begin{align}
\int\!\! d^3x'\gradv\bigg(\!\gradv\cdot\bigg(\frac{[\v J]}{R}\bigg)\!\bigg) &= -\frac{4\pi}{3}\;\v J + \!\int\! d^3x'\bigg(\!\frac{3\hat{\v R}(\hat{\v R}\!\cdot\! [\v J]) - [\v J]}{R^3}
+ \frac{3\hat{\v R}(\hat{\v R} \cdot [\partial \v J/\partial t']) - [\partial \v J/\partial t']}{R^2 c}\nonumber\\
&{}\quad + \frac{\hat{\v R}(\hat{\v R}\cdot[\partial^2 \v J/\partial t'^2])}{Rc^2} \bigg).\label{a18}
\end{align}
From Eqs.~\eqref{a4} and \eqref{a18} we obtain Eq.~\eqref{7}.

\section{Derivation of Eq.~\eqref{43}}

By applying the general identity
\begin{equation}
\frac{[\gradv'\times \v F]}{R}=\gradv\times\left(\frac{[\v F]}{R}\right)+\gradv'\times\left(\frac{[\v F]}{R}\right),\label{b1}
\end{equation}
we can obtain
\begin{equation}
\int\! d^3x'\frac{[\gradv' \times (\gradv' \times \v J)]}{R} =\!\int\! d^3x'\gradv\!\times\!\bigg(\!\frac{[\gradv' \times \v J]}{R}\!\bigg) + \int\! d^3x'\gradv' \times \bigg(\!\frac{[\gradv' \times \v J]}{R}\!\bigg).\label{b2}
\end{equation}
The second integral can be transformed into a surface integral which vanishes at infinity. By applying Eq.~\eqref{b1} again to the first term in Eq.~\eqref{b2}, we obtain
\begin{equation}
\int\! d^3x'\gradv\!\times \bigg(\!\frac{[\gradv' \times \v J]}{R}\!\bigg)=\int\! d^3x'\gradv \times\bigg(\!\gradv \times \bigg(\frac{[\v J]}{R}\bigg)\!\bigg)+\gradv \times\!\int d^3x'\gradv' \times\!\bigg(\frac{[\v J]}{R}\!\bigg).\label{b3}
\end{equation}
The last integral can be transformed into a surface integral and vanishes. From Eqs.~\eqref{b2} and \eqref{b3} we obtain
\begin{equation}
\int\! d^3x'\frac{[\gradv'\!\times\!(\gradv' \times \v J)]}{R} =\!\int\! d^3x'\gradv\!\times\bigg(\!\gradv \!\times\!\bigg(\frac{[\v J]}{R}\!\bigg) \!\bigg).\label{b4}
\end{equation}
In contrast, the identity\cite{23}
\begin{equation}
\Box^2\bigg(\frac{[\v J]}{R}\bigg)= -4\pi[\v J]\delta(\v x-\v x'),\label{b5}
\end{equation}
can be expressed as
\begin{equation}
\gradv \times\bigg(\!\gradv \times\!\bigg(\frac{[\v J]}{R}\bigg)\!\bigg)= 4\pi[\v J]\delta(\v x-\v x')-\frac{[\partial^2 \v J/\partial t'^2]}{Rc^2}
+\gradv\bigg(\!\gradv\cdot\bigg(\!\frac{[\v J]}{R}\bigg)\!\bigg).\label{b6}
\end{equation}
If Eq.~\eqref{a5} is inserted into Eq.~\eqref{b6}, we find
\begin{align}
\gradv \!\times\!\bigg(\!\gradv \!\times \!\bigg(\frac{[\v J]}{R}\bigg)\!\bigg)&\!=\!\frac{3\hat{\v R}(\hat{\v R} \cdot [\v J]) - [\v J]}{R^3}+ \frac{3\hat{\v R}(\hat{\v R} \cdot [\partial \v J/\partial t]) - [\partial \v J/\partial t]}{R^2 c} + \frac{\hat{\v R} \times (\hat{\v R} \times [\partial^2 \v J/\partial t^2])}{Rc^2}\nonumber\\
&{}\quad +\frac{8\pi}{3}[\v J]\delta(\v x-\v x').\label{b7}
\end{align}
If we use Eqs.~\eqref{b4} and \eqref{b7}, we obtain
\begin{align}
\int\! d^3x'\frac{[\gradv' \times (\gradv' \times \v J)]}{R} & = \frac{8\pi}{3}\v J + \int\! d^3x'\bigg(\frac{3\hat{\v R}(\hat{\v R} \cdot [\v J]) - [\v J]}{R^3}+ \frac{3\hat{\v R}(\hat{\v R} \cdot [\partial \v J/\partial t]) - [\partial \v J/\partial t]}{R^2 c}\nonumber\\
&{}\quad + \frac{\hat{\v R} \times (\hat{\v R} \times [\partial^2 \v J/\partial t^2])}{Rc^2}\bigg).\label{b8}
\end{align}
Substitution of Eq.~\eqref{b8} into Eq.~\eqref{41} gives Eq.~\eqref{43}.

\end{document}